**Title:**

A Generalizable Deep Learning System for Cardiac MRI

**Authors List:**

Rohan Shad[1*], Cyril Zakka[2], Dhamanpreet Kaur[2], Mrudang Mathur[2], Robyn Fong[2], Joseph Cho[2], Ross Warren Filice[3], John Mongan[4], Kimberly Kallianos [4], Nishith Khandwala[5], David Eng[5], Matthew Leipzig[2], Walter R. Witschey[6], Alejandro de Feria[7], Victor A. Ferrari[7], Euan A. Ashley[8], Michael A. Acker[1], Curtis Langlotz[9], William Hiesinger[2*].

**Affiliations:**

1. Division of Cardiovascular Surgery, Department of Surgery, University of Pennsylvania
2. Department of Cardiothoracic Surgery, Stanford University
3. Department of Radiology, Medstar Georgetown University Hospital
4. Department of Radiology and Biomedical Imaging, University of California, San Francisco
5. Bunkerhill Health, San Francisco
6. Department of Radiology, University of Pennsylvania
7. Division of Cardiovascular Medicine, Department of Medicine, University of Pennsylvania
8. Division of Cardiovascular Medicine, Department of Medicine, Genetics, and Biomedical Data Science, Stanford University
9. Department of Radiology, Medicine, and Biomedical Data Science, Stanford University

(*) – Corresponding Author

Rohan Shad, MD
rohan.shad@pennmedicine.upenn.edu

William Hiesinger, MD
willhies@stanford.edu

**Abstract**

Cardiac MRI allows for a comprehensive assessment of myocardial structure, function, and tissue characteristics. Here we describe a foundational vision system for cardiac MRI, capable of representing the breadth of human cardiovascular disease and health. Our deep learning model is trained via self-supervised contrastive learning, by which visual concepts in cine-sequence cardiac MRI scans are learned from the raw text of the accompanying radiology reports. We train and evaluate our model on data from four large academic clinical institutions in the United States. We additionally showcase the performance of our models on the UK BioBank, and two additional publicly available external datasets. We explore emergent capabilities of our system, and demonstrate remarkable performance across a range of tasks; including the problem of left ventricular ejection fraction regression, and the diagnosis of 39 different conditions such as cardiac amyloidosis and hypertrophic cardiomyopathy. We show that our deep learning system is capable of not only contextualizing the staggering complexity of human cardiovascular disease, but can be directed towards clinical problems of interest yielding impressive, clinical grade diagnostic accuracy with a fraction of the training data typically required for such tasks.

**Main text:**

## Introduction:

Offering unparalleled diagnostic clarity, Cardiac Magnetic Resonance Imaging (CMR or Cardiac MRI) is the reference standard for assessing cardiac anatomy and function.[1,2] Depending on the technique used, it enables clinicians to capture videographic sequences of cardiac and valvular motion, quantify scarring or tissue edema within the cardiac musculature, and identify regions of poor perfusion – all without exposure to ionizing radiation.[2] Despite this wealth of data available, deep learning systems capable of learning high quality representations of human cardiac disease from CMR have not yet been described.

Deep learning has shown incredible promise in the diagnosis of cardiovascular disease with EKG, retinal scans, and echocardiography.[3–7] Traditionally, these networks are trained to detect a handful of pre-defined and curated 'disease' conditions in the background of negative 'normal' cases. While superficially successful, many of these systems fail when tested on real world clinical data that is often heterogenous with numerous concomitant abnormal findings.[8] Patients with inherited cardiomyopathies, for example, may present with severe valvular disease; those with evidence of ventricular thrombus on the other hand, may have severe heart failure from remote ischemic insults. With the traditional supervised deep learning approach, it remains challenging to contextualize this diversity of disease presentation from CMR scans, with parameters learned for one problem rarely generalizing to others.[9] These systems, therefore, must be re-trained from scratch for every new clinical task of interest, requiring thousands of training examples each time. Unlike human clinicians, deep learning models do not have a baseline fund of clinical and pathophysiological understanding over which learning specific tasks can be accelerated. Ultimately this has restricted research in the field to tasks that either automate objective structural measurements or to diseases that are simply more prevalent.[10]

Here we describe a transformer-based vision system that learns complex pathophysiological visual representations from a large multi-institutional dataset of 19,041 CMR scans, guided by natural language supervision from the text reports accompanying each CMR study. We use a large language model to help 'teach' a vision network to generate meaningful low-dimensional representations of CMR studies, by showing examples of how radiologists describe what they see while drafting their reports. We describe how these models group together patients with similar pathophysiological and even socio-demographic characteristics with no explicit supervision for these tasks. We finetune this deep learning system on smaller expert labelled datasets for a range of clinically relevant tasks ranging from left ventricular ejection fraction estimation, to the detection of 39 different cardiovascular conditions. We validate our system on multiple external datasets from distinct geographical sites and health systems. The result is a generalizable and data-efficient CMR deep learning system capable of representing the breadth of cardiovascular disease and health.  Fig 1. details an overview of the project.

## Cardiac Magnetic Resonance Imaging data:

The inputs to our networks are SSFP (Steady State Free Precession) cine-sequences taken along multiple different cardiac view planes. Typically, SSFP sequences are characterized by a high blood signal intensity, with lower intensity myocardial signal.[11,12] This technique allows for the acquisition of high contrast dynamic motion scans of the heart during a breath-hold timed with EKG gating to capture images between heart beats. These cine-sequence videos are captured along various spatial view planes as described below. Our data is sourced from a heterogenous sample population with substantial variability in acquisition and scanner vendors, with training data sourced from Stanford Medicine, MedStar, and UCSF. Additional data from the UK BioBank, Kaggle Grand Data Challenge, ACDC Dataset, and the University of Pennsylvania were sourced to rigorously assess model generalizability. Specific descriptions of each cohort are provided in the Methods and Supplementary Fig 1.

Different cardiac structures may be visualized better in one view plane over another. As a result, findings indicative of pathology may only be visible in a specific MRI view plane. This is conceptually analogous to the challenges faced in deep learning for histopathology, where only a fraction of the biopsied tissue might contain findings diagnostic of disease.[13] We limit ourselves to short axis stacks (SAX), four-chamber (4CH), three-chamber (3CH), and two-chamber (2CH) views for this work as these are most consistently captured irrespective of query pathology. Demographic data available for certain subsets of the data are detailed in Supplementary Table 1.

## Pretraining framework and evaluation of low-dimensional representations:

The traditional approach for deep learning with cardiovascular imaging has been to assign some label to each imaging scan via tedious manual annotation. Large scale labelling of training data requires extensive domain knowledge and clinical expertise, but ultimately constrains the models to features explicitly labelled in the dataset. Models trained this way require thousands of examples for adequate performance, and are unable to account for findings outside the scope of what they were trained to identify. Recent advances in self-supervised deep learning have shown promise in reducing this reliance on vast quantities of expert labelled data across a wide range of modalities ranging from pathology slides to chest X-rays.[14–21] Yuhao et. al describe a framework wherein unstructured text could be used as a method of self-supervision for a Chest X-Ray deep learning system.[16] In this method of contrastive learning, two neural networks are used to produce a pair of low-dimensional representations for each pair of contextually related inputs from two separate modalities.[22] We extend these concepts to the spatiotemporal and multi-view problem of Cardiac MRI. During the pre-training process, the networks are trained to match true pairs of text report and MRI scans by optimizing for a contrastive objective.[22] Since the release of our initial preprint, others have reported promising results using similar contrastive-learning methods to train generalizable "foundation models" for computed tomography (CT) and echocardiography[23–25].

Visual features of a particular disease are iteratively associated with textual cues, with a training process guided by a rich unstructured description of disease created by clinicians as part of routine clinical workflow. We use an implementation of a multi-scale vision transformer (mVIT) for the CMR vision encoder, and a BERT (bidirectional encoder representations from transformers) text transformer for the MRI report encoder.[26,27] We pretrain and validate on 14073 cardiac MRI scans (12707 unique patients) from Stanford, UCSF, and MedStar. Specific implementation details are described in the Methods.

We plot the embeddings generated by the vision encoder on the validation set using standard dimensional reduction algorithms (Methods), demonstrating the progressive emergence of local and global structure in the two-dimensional projections of these embeddings as training progresses (Fig 1.). Once training is complete, the network is frozen and deployed onto a publicly available external dataset of cine-CMR sequences sourced from France (ACDC Dataset).[28] We plot the embeddings for each study in the ACDC Dataset onto two dimensions, and confirm that compared to a baseline network of the same architecture, our vision system is able to separate different disease conditions with remarkable consistency (Supplementary Fig. 4). This is despite the lack of any directed supervision during the training process in separating conditions such as hypertrophic cardiomyopathy from dilated cardiomyopathy or right ventricular dysfunction.

## Validation on the UK BioBank:

We sourced 159,883 cine-CMR scans representing 45,623 participants from the UK Biobank, and assessed the performance of large-scale contrastive pretraining on a dataset representing a relatively healthy participant population from a geographically distinct continent. We hypothesized that if useful representations are learned by our deep learning system during the pre-training phase, it should be trivial to separate participants based on features indicative of disease without any additional supervised training. We use our contrastive pre-trained network and freeze the weights to prevent any additional learning, following which we pass each available cine-CMR sequence through the network. The network generates a low-dimensional embedding of each input. We process each view separately, and store the embeddings generated for further processing. As with the experiments on the ACDC dataset, we repeated the experiment with an identical network with weights initialized from a training run on the Kinetics-400 action recognition dataset – a video dataset of natural scenery and activities.[29]

We elected to use the unsupervised t-SNE algorithm to dimensionally reduce the embeddings generated for each view, and visualize them in two-dimensions.[30] We expected the networks at baseline to at least be capable of generating embeddings of sufficient quality to separate different views when visualized with t-SNE, as they are obviously distinct even to the untrained eye. In Fig 2 (left) we show the unsupervised clustering of embeddings produced by our contrastive pre-trained network in 2D t-SNE space, for all 45,623 UK Biobank participants. We find that the pre-trained network is essentially view-invariant. Key to driving invariance based on view-planes was to ensure all view planes from the same study are aligned with a text-embedding from the same text-report during pre-training. This allows for attention to be placed instead on features of importance highlighted in the text. We see a relatively dense cluster of patients with ejection fractions < 35%. Surprisingly, without any explicit instruction, we find that contrastive pre-training allowed for sharp delineation of sex, and separation by age. It is likely that the presence of textual information containing phrases such as 'ADULT FEMALE' within some of the MRI reports allowed the vision networks to learn features that identify demographic characteristics. Importantly, all of these findings are seen with the networks entirely frozen, with no additional learning or model parameter updates possible. Comparing with similar experiments using the Kinetics-400 pre-trained baseline is instructive in demonstrating the lack of any such biomedical latent knowledge via traditional training methods (Fig 2; right). No distinct clusters beyond those of the major MRI view planes studied are seen without our contrastive pretraining methods.

Self-attention based aggregation of embeddings from multiple CMR Views:

We perform a series of experiments to define the impact of contrastive pretraining when these models are tasked to a variety of clinically relevant regression and classification problems. Each view in a cardiac MRI study may be of varying diagnostic utility, depending on the presenting pathology. To mimic the approach of clinicians reporting cardiac MRI scans, a secondary neural network was used to aggregate useful information from each view plane to generate a final scan level output (Methods). As a result, our models use all available views (4CH, 2CH, 3CH, and SAX) while analyzing each patient exam. We treat the presence of multiple MRI view planes as a weakly labelled multi-instance learning problem. Videos from each view plane are processed via our pre-trained video transformer encoder, and the resulting embeddings are fed into a self-attention based multi-instance learning module. The self-attention module is trained to identify relevant features across each available view plane, independent of the actual number of views supplied.[31] The relative contribution of one view over another towards the final output is calculated via a learned weighted average of the output embeddings. This assigns importance to different views depending on the pathological features present in them. This is conceptually identical to assigning higher importance to a selection of zoomed in patches of histopathology whole slide images that contain features of malignancy, in the background of a large number of non-diagnostic patches.[13,31]

Automated Estimation of Left Ventricular Ejection Fraction:

We define the impact of contrastive pretraining on problem of predicting left ventricular ejection fraction (LVEF) – a commonly calculated metric of cardiac function on two external datasets – the UK BioBank referenced above, and a publicly available CMR dataset (Kaggle) for this problem. Deep learning derived automated LVEF measurements in echocardiography and cardiac MRI are typically based on segmentation models or hybrid regression-segmentation systems.[4,28,32,33] Such networks are trained to outline left ventricular chambers during end-systole and end-diastole, and as such have no further inherent understanding of the disease space. Nonetheless, these networks perform exceptionally well in the task of LVEF estimation as they replicate the measurement workflow of routine clinical practice. Bai *et. al* for instance, report a mean absolute error (mae) of 3.2 for the task of predicting LVEF in a subset of the UK Biobank participants using short-axis sequences.[34] Two recent papers released while our manuscript was in review are notable for their technical advance in the area of deep-learning powered LVEF estimation: Fu *et al* describe CineMA, a masked-autoencoder system utilizing a hybrid-segmentation supervision; and Zhang *et al* who describe masked autoencoders combined with a contrastive objective.[35,36] While both achieve competitive performance on LVEF estimation in the UK BioBank (CineMA with a mae of 3.34% and Zhang *et al.* with a mae of 2.95%), critically, these systems were both pre-trained and finetuned on the UK BioBank itself making direct comparisons of generalizability challenging.

Prospective studies have shown that there is considerable variability in institutional and dataset specific protocols for calculation of certain metrics, clinicians can typically be expected to make estimates of LVEF within Bland-Altman limits of -12% to +12%.[37,38] With our pretrained vision encoder frozen barring the last linear layer, we finetune the network on 34488 unique CMR scans from the UK BioBank for the task of predicting LVEF. The vision encoder can no longer learn dataset specific features with this approach, and must rely on learned representational abilities from prior pre-training. To incorporate information from each available view to produce a 'patient level' prediction of LVEF we use a multi-instance self-attention regression head as described above, and input cine-CMR sequences from all available views without any additional quality control steps (Methods). On a hold-out test subset of UK BioBank participants using our approach, we report a mean absolute error (mae) of 3.344 (sd 3.615), with Bland Altman limits of agreement -9.91% to +9.61%. At baseline, using a traditionally trained deep learning system of the same model architecture as our contrastive pre-trained models, we report mae of 4.603 (sd 4.409), with Bland Altman limits of agreement -12.15% to +12.8%.

These metrics rival those of hand-crafted deep learning systems to calculate LVEF using segmentation masks on manually selected end-systolic and end-diastolic frames, and are well within error limits of clinicians following strict annotation protocols.[35,37] We freeze this finetuned model and evaluate it on cardiac MRI scans from a publicly available dataset, where patients were recruited from hospital systems based in the United States (Kaggle Data Science Bowl)[39]. The Kaggle external dataset contains a larger proportion of patients with diseased hearts, and additionally utilizes a slightly different scanning protocol and method for determination of ground truth LVEF labels. On this dataset we report a mae of 6.880 (sd 5.309); with Bland Altman limits of agreement -18.7% to + 8.03%. Given the differences in annotation methods compared to data from the UK BioBank, the frozen finetuned network showed a modest systematic underprediction of LVEF by 5.36% (95% CI: 4.86 – 5.87)(Fig 3b). Diagnostic plots for test-set results and additional results with improvements on the Kaggle dataset with bias correction are available in Supplementary Fig. 7 and 8 (final bias corrected mae 4.861), showing superior performance in the high and low EF range with smaller residual errors when models are initialized from contrastive pre-trained weights. Finally, we manually review CMR studies for model predictions with the largest absolute errors, and find the majority of these stem from incorrect ground truth labels or artifact / noise degraded images (Supplementary Figure 9, and Table 4).

Nevertheless, if these predicted LVEF values were to be used for identifying patients with HFrEF (Heart Failure with reduced ejection fraction <40%),[40] the AUC on the UK BioBank test set would be 0.880 (95%CI: 0.835-0.925; n=4259 scans), and 0.949 (95% CI 0.907-0.990; n=679 scans) on the external Kaggle dataset (Fig 3c). This is significantly more performant compared to our baseline, where we finetune the exact same network architecture from a Kinetics-400 checkpoint. Baseline performance on the UK BioBank test set would be 0.751 (95% CI: 0.692-0.811; $p < 0.001$) and for the Kaggle dataset 0.766 (95% CI: 0.697-0.836; $p < 0.001$) (Supplementary Table 5). This is in line with recent reports of poor generalizability with models trained via more traditional methods on the UK BioBank.[37] We find that using our contrastive pre-trained vision encoders allow for impressively low error rates when finetuning on just 1% of the data available in the UK BioBank dataset (344 scans), far surpassing our baselines (Fig 3a; Supplementary Table 3). We also find that unfreezing the vision encoder allowing for parameters to be overwritten (transfer learning) with new learned weights led to a marked decline in performance across the board. To further explore the relationship of pre-training quality on downstream performance on the task of LVEF prediction, we finetune models from checkpoints taken at equally spaced intervals during the pre-training process. We find that downstream validation performance continues to increase with decreasing pre-training loss in a non-monotonic fashion (Fig 3d). Finally, we studied the scan-rescan variability of LVEF predictions generated by our models for participants with more than one available CMR study acquired at two distinct time periods (n = 311; test-set). The mean variance was found to be 5.98% (sd 1.53%), with Bland Altman limits of agreement between scans of -6 to +6%. These figures exceed previous benchmarks of clinical expert level performance in prospective trials (Supplementary Table 9 and Supplementary Fig 14).[37]

### Efficiently diagnosing disease from Cardiac MRI:

Cardiac MRI is routinely used to assess patients with left ventricular dysfunction, congenital heart disease, valvular disease, and various cardiomyopathies.[41] While morphological features seen on cine-sequences may be sufficient to diagnose diseases such as hypertrophic cardiomyopathy, others such as cardiac amyloid or the presence of an intracardiac thrombus typically require contrast enhanced sequences. Prior work in automated deep learning-based disease diagnosis has relied on segmentation of cardiac chambers or specific patterns of scar, using limited datasets that lacked the comprehensive inclusion of a wide range of cardiac conditions.[42,43]

We created a labelled dataset of 4301 unique patients from Stanford, Medstar, and UCSF distinct from the pre-training dataset for 39 cardiovascular conditions using reports generated by expert radiologists as part of routine clinical practice. We intentionally include disease conditions that typically require the evaluation of contrast enhanced images or T1/T2 sequences for accurate clinical diagnosis. Furthermore, we included certain labels that are more subjective in nature (eg: 'severe' valvular disease). These datasets were unenriched, uncurated, and reflect the real-world clinical prevalence at the various clinical sites (Supplementary Fig 10 and 11).

With our pretrained vision encoder frozen barring the last linear layer, we finetune the network to predict the binary classification label for each disease task on a subset of 2145 unique patients (2414 scans). We generate a 'patient level' prediction using a multi-instance self-attention classification head, and input cine-CMR sequences from all available views and slices without additional quality control steps (Methods). Models were validated and tested on separate splits of 1073 and 1083 patients respectively. We show significant absolute improvements in AUROC across most clinical tasks when finetuning from our pre-trained models, compared to baselines finetuned from a Kinetics-400 pre-trained checkpoint (Fig 4a, Supplementary Tables 6 and 7; Supplementary Fig 12). Furthermore, attention maps plotted from the multi-instance self-attention module show that embeddings from different views are preferentially weighted depending on the clinical task at hand (Fig 4c).

### External validation on the Penn Cardiac MRI dataset

We follow a similar data labelling technique as described above for CMR scans from the University of Pennsylvania. Final model checkpoints for each disease task from the experiments above were frozen, and subsequently used for testing on the Penn dataset. No additional fine-tuning was performed on this data. Testing was performed in series for each disease label on either a single Nvidia A40 GPU or Nvidia B200 GPU, taking about 10 minutes each, for a total of 2070 scans from 2033 unique patients. As before, all available view planes (2CH, 3CH, 4CH, SAX slices) were used with no additional quality control or filtering steps. On this external test dataset, we show robust performance for disease conditions such as Amyloidosis, Hypertrophic Cardiomyopathy and Tetralogy, but significant degradation for labels such as Ventricular Septal Defects and Sarcoidosis. AUROC curves and confidence intervals are shown in (Fig 4b; Supplementary Table 8, Supplementary Fig 13).

On review of institutional CMR sequence protocols, a possible contributing factor for our findings is that at the University of Pennsylvania, the standard practice is to administer contrast prior to the acquisition of all cine-sequences – a technique employed to reduce scan times. This was not the case at Stanford, MedStar, and UCSF where cine sequences are acquired before contrast is administered. This represents a dramatic shift in the underlying pixel-data distribution. Despite these differences in image acquisition protocols, the robust performance on several tasks suggest that our models rely on structural and dynamic motion features to identify certain disease states, rather than patterns of pixel intensities alone that are not robust to timing of contrast.

### Diagnostic performance strengths and limitations

Despite the absence of contrast enhanced images in the inputs, our systems excelled at the detection of several cardiomyopathies on the test sets: Internal test-AUCs for Amyloidosis was 0.921 (95% CI 0.879 – 0.963), Hypertrophic Cardiomyopathy 0.91 (95% CI 0.872 – 0.948), and Dilated Cardiomyopathy 0.867 (95% CI 0.843 – 0.890). However, performance on tasks involving the detection of intracardiac thrombi (AUC 0.744; 95% CI 0.653 – 0.836) or certain cardiac tumors (AUC 0.649; 95% CI 0.535 – 0.764) remained marginal. Differentiating tumor from thrombus, or diagnosing Myocarditis can be particularly challenging for clinicians without contrast enhancement or T2 imaging [44–46]. The Lake Louise imaging criteria used for the diagnosis of Myocarditis in clinical practice, for instance, relies on T1, T2, or LGE sequences.[47] Our results indicate that cine-sequences alone are unlikely to provide enough signal to reliably diagnose such conditions despite the alignment with textual information in the reports. Similarly, the diagnosis of Arrhythmogenic Right Ventricular Cardiomyopathy (ARVC) relies on a Clinical Task Force Criteria (TFC) scoring system: a combination of clinical symptoms, medical and family history, EKG findings, and imaging findings.[48] Our model performance returns an AUC of 0.652 (95% CI 0.487 − 0.818) for ARVC, indicating the challenges of diagnosing conditions - that in routine clinical practice - rely on a comprehensive evaluation of the patient beyond what images alone can provide.

Valvular heart disease, similarly, was challenging to accurately grade for our models, though this is likely the result of poorly defined severity labels in the MRI reports themselves. Severe Mitral Regurgitation (AUC 0.753; 95% CI 0.716 – 0.79) and Severe Aortic Stenosis (AUC 0.706; 95% CI 0.622 – 0.79) are characterized by either turbulent regurgitation jets or high velocity stenotic jets respectively. Detecting bicuspid valves on the other hand requires acquisition of images in the plane of the aortic valve to truly define the location and orientation of fused aortic valve leaflets. While these features are sometimes visible on cine sequences, accurately quantifying regurgitant volumes and the severity of valvular heart disease requires additional post-processing of phase contrast sequences in specific anatomical planes or 4D-flow scans that are seldom performed as part of standard CMR scanning protocols.[49]

Our models performed well in detecting anomalous anatomical configurations such as Tetralogy of Fallot (AUC 0.966; 95% CI 0.955 – 0.977), Ventricular Septal Defects (AUC 0.911; 95% CI 0.884 – 0.938), and changes in ventricular mass. The performance across tasks appears unrelated to the underlying prevalence of the disease labels themselves, with excellent performance for diseases such as Amyloid (prevalence ~ 1%), and worse performance for those such as RA Dilation (prevalence 13%). Results are summarized in Supplementary Table 6, and Fig 4. In clinical practice, Cardiac MRI is often used to establish differential diagnoses, for example interrogating the underlying etiologies of individuals presenting with heart failure. To illustrate this, we analyze a subset of patients with the clinical label of "Non-Ischemic Cardiomyopathy" (n = 242) in our internal test dataset. This broadly includes cardiomyopathies associated with infiltrative diseases, genetic, hypertrophic or dilated cardiomyopathies, and myocarditis among others. Such subgroup analyses demonstrate conditional model performance in clinically challenging contexts, for instance: Certain non-ischemic cardiomyopathies share overlapping phenotypic features of left-ventricular hypertrophy (sarcoidosis, amyloidosis, and hypertrophic cardiomyopathy). This makes discerning the underlying etiology challenging for models initially designed to detect disease in the background of phenotypically very distinct negative classes. Nonetheless, our models demonstrate impressive performance in this subset for Amyloidosis (AUC 0.870; 95% CI 0.675 – 0.957), Dilated Cardiomyopathy (AUC 0.879; 95% CI 0.836 – 0.923), and Hypertrophic cardiomyopathy (AUC 0.925; 95% CI 0.873 – 0.977) (Supplementary Table 18).

Mirroring findings on the UK Biobank LVEF% regression task, there is significant performance degradation when attempting to transfer learn with all vision encoder parameters unfrozen, suggesting the fundamental importance of parameters learned via contrastive pre-training. Finally, we find on scan-rescan testing that the variance between predicted probabilities of patients with more than one scan is on average less than 3% in both the internal test-set and the external UPenn test datasets (Supplementary Tables 10 and 11).

### Clinical applications and future directions

The core vision system remains unchanged in our proposed framework: We pre-train one vision network capable of contextualizing and representing the features important for multiple actionable downstream clinical tasks of interest. Our principal contribution is the development of a system that is readily adaptable with minimal additional finetuning data, to a wide range of clinical tasks. As a result, there are numerous immediate clinical applications. We anticipate our models will enable rapid and expert-level diagnosis of relatively rare complex cardiomyopathies, democratizing expertise beyond the confines of large academic centers of excellence. This is particularly relevant for conditions such as Hypertrophic Cardiomyopathy, and Cardiac Amyloidosis that are often misdiagnosed or under-diagnosed in the community.[50,51] Our models may serve as a specialist level triage, opportunistically labelling patients with a high predicted probability of having disease as soon as scans are completed for physicians to review. At a systems level, our models can help automatically triaging and notifying clinical services, ensuring flagged patients are formally evaluated by subspeciality cardiology teams. This is especially important for cardiomyopathies that can be effectively managed with novel pharmacological agents with significant reduction in mortality and morbidity.[52,53]

Our methods, furthermore, make possible for deep learning systems to be adapted towards identifying imaging features that are predictive of a histopathological ground truth – narrowing the modality gap between radiological and histopathological diagnoses for diseases such as amyloidosis, sarcoidosis, post-heart transplant rejection, and cardiac malignancies. This has the potential of reducing our reliance on invasive testing for obtaining actionable diagnostic information, dramatically reducing the risk of iatrogenic injury.[54] We envision our models could be deployed as a clinical decision aide, potentially allowing us to reserve invasive endomyocardial biopsies for patients who have indeterminate results via our models.

We recently presented work on our pre-trained models in screening the UK BioBank population for Hypertrophic Cardiomyopathy.[55] CMR studies acquired as part of the UK BioBank are usually not reviewed by an imaging cardiologist unless glaring abnormalities are noted, likely contributing to the 5-10x lower reported prevalence compared to the general population.[56] Given the overwhelming majority of normal studies, cases exceeding a probability threshold of 95% – correlating to the top 0.5% of cases (expected specificity of 97% and sensitivity of 60%) – were screened in for manual reading. With this approach, our group screened through over 40,000 CMR scans, and identified 200 scans for manual physician over-read. From this high probability subpopulation, we confirmed 112 new individuals with imaging features of HCM. Our models thus have significant potential for clinical expert-grade phenotypic refinement of large research datasets.

Traditional uses for deep learning systems for CMR are in the automation of otherwise hand-measured metrics such as LVEF, chamber volumes, or the estimation of myocardial thickness. Our approach is to use the same pre-trained vision system, obviating the need for labour intensive data annotation for each new structure or target of interest. While segmentation-based tools are able to achieve similar results, they are critically limited in needing ground-up retraining for every new metric of interest, with significant human effort required for each new task. Our modular approach of swapping regression or classifier heads depending on the task of interest allows for rapid repurposing of representations learned by our core vision encoder.

Finally, we show that our models are capable of intriguing emergent properties out-of-the-box, including the ability to separate sex, age, and certain disease phenotypes without additional finetuning or instruction. While traditional approaches are limited to standard structural measurements of the heart, these properties of our models make them powerful tools for efforts directed at phenotypic refinement or genetic discovery in cardiovascular disease in large population-scale data repositories.[57–62]

## Discussion:

The methodology for our foundational deep learning CMR model incorporates contrastive pre-training over a joint embedding space. Visual features are extracted from scans, and text features are extracted from reports generated by clinical experts as part of routine care. Unlike prior work on Chest X-Rays where an image is taken from a largely consistent anatomical view, for CMR studies, the inputs are a series of videos acquired along multiple distinct anatomical view planes. Even within a particular view plane, there may be multiple unique parallel videos acquired (eg. cine-bSSFP sequences at the base, mid-papillary, and apical cross-sections of the heart in the SAX view). The contrastive pre-training routine was thus designed to maximize agreement with shared text-embeddings across all available views present within a single study. We hypothesize that this solves two interdependent and critical problems: First, that traditional supervised learning methods often exacerbate the problem of 'shortcut learning', wherein deep learning systems approximate simple decision rules that allow for networks to perform well for a narrow task.[9] Second, the parameters learned for a specific and narrow clinical problem rarely transfer to new tasks. This demands time and labour-intensive labelling of large amounts of data for each clinical task of interest.

By pre-training and subsequently finetuning towards specific problems, we show that with remarkable consistency, our networks perform superior to baseline methods. We show that across numerous unrelated tasks, superior performance can be achieved with up to two orders of magnitude less data. This data efficiency is a critical advance for the development of finetuned models for the diagnosis and characterization of complex inherited cardiomyopathies, where expert adjudicated registries with paired imaging data remain scarce.[63,64]

While pre-training itself is computationally expensive, finetuning on each task can be achieved within a few hours on consumer grade GPUs. In inference mode with no additional hardware specific optimizations, our models can process an entire CMR study with multiple views, sequences, and anatomical slices in under 400ms. With additional optimization and quantization, these models can readily run on resource efficient embedded systems. This allows for clinical deployment strategies ranging from local on-device inference to light-weight cloud containerized instances that interface with hospital PACS (Picture Archiving and Communication System) servers.[65] Furthermore, the pretraining framework can be extended to data sources beyond linked CMR reports, including text-reports from other imaging modalities (eg: echocardiography, cardiac catheterization reports) or histopathological slide features that may further expand the emergent capabilities of vision encoders seen in this work.

Large language models have grown immensely in capabilities and in size. Our implementation of BERT has 110 million trainable parameters and just over a 30,000 token vocabulary, whereas contemporary large language models boast up 80 billion parameters and over a trillion tokens. Contrastive pre-training makes use of embeddings produced by the text-encoder (a feature of architectures such as BERT). The "decoder-only" architectures of most contemporary large language models preclude direct extraction of embeddings needed to train on a joint-representation space. Recent advances offer alternatives, but are too computationally expensive to be viable in their current forms.[66]

Another limitation of our current work is the reliance on cine-SSFP sequences alone. While the reports describe findings from gadolinium enhanced imaging, the networks do not make use of LGE (Late Gadolinium Enhanced) sequences as inputs. This is largely because LGE sequences are often captured as images rather than dynamic-motion videos, necessitating a separate and parallel image-based neural network to be built into the current framework. Future research will incorporate non-standard view-planes along with T1, T2, LGE, and perfusion scans via separate image encoders or dimensionally agnostic vision encoders capable of handling both image and video data. Despite this, our models consistently perform well on tasks with just cine-SSFP sequences in situations where clinicians typically require additional imaging data. LGE, for instance, is routinely used by clinicians for diagnosing Amyloidosis; similarly LGE along with T1, and ECV (extracellular volume) maps are often used in diagnosing Hypertrophic cardiomyopathy.[67–69]

We hope our work accelerates research in this field, making it feasible to apply deep learning techniques to clinical areas of interest on datasets traditionally deemed too small or niche. Finetuning on smaller, expert labelled datasets where ground truth is established via either expert consensus or histopathological reads of myocardial biopsies are exciting avenues of future research. Keeping this in mind, our pre-trained checkpoints are made freely available for academic use. Our results are an important step in the evolution of deep learning for cardiac MRI. In summary, we describe a generalizable self-supervised deep learning system for cardiac MRI capable of representing the breadth of human cardiovascular disease. This work lays the foundation for prospective clinical grade applications in disease diagnosis, with immediate research applications in phenotyping and genetic discovery of cardiovascular disease using cardiac MRI.

## Methods

### Ethics Statement

Data collection and research began following approval and waiver of consent by the Institutional review board of Stanford University (Protocol #60342; March 2021). Additional data from the University of Pennsylvania was sourced after retrospective collection was deemed IRB exempt by the University of Pennsylvania Health System (Protocol #852332; November 2022).

### Computational hardware and software

MRI DICOM data were pre-processed on siloed HIPAA certified n2-instances on the Stanford Nero - Google Cloud platform. Specifically, we used an 8-core virtual machine with 52GB of memory, and 6TB of attached solid state storage. Data from the UK Biobank were pre-processed on the Stanford Sherlock High Performance Computing Cluster, using 24 CPU cores (Intel Xeon Gold 5118; 2.30Ghz). Anonymized reports were tokenized on a local encrypted desktop using 48 CPU cores (AMD Threadripper; Lambda Computers). All models were trained on the Stanford Sherlock High Performance Computing Cluster using servers with 4x Nvidia A100 GPUs each with either 40GB or 80GB VRAM, and 64 CPU Cores (AMD Epyc). External validation on data from the University of Pennsylvania was performed on the Penn CUBIC High Performance Computing Cluster on a single Nvidia A40 GPU with 10 CPU cores. Additional external tests took place on the Penn Advanced Research Computing Center (PARCC) Betty cluster, on a single Nvidia Blackwell B200 GPU with 10 CPU cores. Hyperparameter optimization experiments were run on servers with a variety of GPU resources (Nvidia V100; 32GB VRAM, Nvidia H100; 80GB VRAM, Nvidia A100; 40GB / 80 GB VRAM, Nvidia P100; 32GB VRAM, Nvidia Blackwell RTX 6000 MaxQ 96GB VRAM). We use the pytorch deep learning library (v1.11.0) and the pytorch lightning framework (v1.8.6).[70] Major python packages used in this work include numpy (v1.21.2), pydicom (v2.0.0), transformers (v4.4.2), and stanza (v1.5.0).

### Datasets:

Specifics of the preprocessing pipelines for both the MRI scans and the free-text reports are detailed in the Dataset pre-processing section of the Supplementary Appendix. Briefly, from each unique MRI study, relevant scans were extracted (4-chamber, 3-chamber, 2-chamber, and short-axis cine sequences) as 4-dimensional arrays and stored within a single hdf5 file. Free text from the reports were segmented into individual sentences using the stanza natural language processing pipeline, and then tokenized using the standard BERT auto-tokenizer and the resulting anonymized numeric arrays were stored in a single indexed json file.[27,71] Across the pre-training datasets, finetuning datasets, external test datasets, and the UK BioBank, we include 65,492 individuals with approximately 550,156 unique videos across different view planes and cross sections.

#### Clinical CMR dataset:

The total clinical CMR dataset comprised of 19122 unique individuals. Cardiac MRI scans were sourced from 17,088 individual patients from a consortium of academic hospital systems based in the United States (Stanford Healthcare, UCSF, Medstar). Cine MRI scans were procured via Bunkerhill Health (San Francisco, CA) as de-identified DICOM files, and associated radiology reports were sourced as a single csv file (IRB Protocol #60342; March 2021). The total pre-training dataset consisted of 293,110 unique 4CH, 3CH, 2CH and SAX videos. The scans were performed as part of routine clinical practice, and reports were generated by board certified physicians with specific expertise in Cardiac MRI. Sequences were acquired on a wide range of scanners including those manufactured by Siemens (Siemens Healthcare, Germany), General Electric (GE, United States), and Philips (Philips Healthcare, Netherlands), resulting in substantial variance in the number of frames per slice, imaging resolution, and reconstruction techniques (Supplementary Table 2). Demographics wherever feasible are detailed in Supplementary Table 1. The data were first separated into pre-training and downstream datasets in an approximate 75:25 split at the patient level. For the pretraining split, we further divided the data into a training and validation set with an approximate 66:33 split. Similarly, for downstream split (intended to be used as a labelled finetuning dataset for clinical tasks of interest) we further divided the data into training, validation, and testing datasets with an approximate 50:25:25 split. We did not selectively exclude patients from this dataset, however a fraction of the dicom files were received as duplicates or were corrupted and were subsequently discarded. Supplementary Fig. 1 details the data splits and enumerates the excluded studies at each stage. Cardiac MRI scans from an additional 2033 individual patients were secured from the University of Pennsylvania Health System (IRB exempt; Protocol #852332; November 2022). These scans were performed as part of routine clinical practice and acquired on scanners manufactured by Siemens and GE. Data from the University of Pennsylvania was used solely for external testing. While rule based automated data labelling techniques have been used in the past, these have been superseded by large language models.[72] Building on our prior work in exploring the zero-shot capabilities of large language models for medical text, we utilized a publicly available large language model (medgemma3; 27 billion parameter variant) to parse free text reports generated as part of routine clinical practice into pre-defined "disease labels" for the disease diagnosis tasks.[73,74] Specific prompts, parameters, performance comparisons vs human annotators, and a selection of random non-curated reports with critique of the deep learning predicted labels are detailed in Supplementary Fig. 5.

### UK Biobank Cardiac MRI cohort

Cine bSSFP-Cardiac MRI sequences from 45,623 participants were sourced from the UK Biobank (Project ID: 71226). SAX sequences were available for 11,005 participants and contain stacks of 8-10 individual slices. One slice was available for each of the 4CH, 3CH, and 2CH scans. This amounted to a total of 257,046 unique videos available for analysis. Sequences in the UK Biobank were acquired on a clinical 1.5 Tesla scanner using a standardized protocol (MAGNETOM Aera, Syngo Platform VD13A, Siemens Healthcare, Germany).[56] As part of this protocol, the vast majority of ventricular volumes and functional metrics are calculated via automated contouring of the ventricular endocardium and epicardium without manual expert quality controls.[56,75] For finetuning and transfer learning experiments to estimate LVEF% we split the UK BioBank dataset into an approximate 80:10:10 split at the participant level into training (n=31693), validation (n=3938), and hold-out test datasets (n=3938).

### Automated Cardiac Diagnosis dataset

The Automated Cardiac Diagnosis Dataset (ACDC) is a publicly available cardiac MRI dataset of 100 patients from the University Hospital of Dijon, France.[28] Each SAX sequence was paired with patient level non-overlapping patient level labels (n=20 each) for hypertrophic cardiomyopathy, previous myocardial infarction, dilated cardiomyopathy, abnormal right ventricles, and normal controls. The scans were acquired on either a 1.5 Tesla (Siemens Area, Siemens Healthcare, Germany) or 3.0 Tesla scanner (Siemens Trio Tim, Siemens Healthcare, Germany) with a conventional SSFP sequence in breath hold and gating.

### Kaggle Data Challenge dataset

The 2015 Kaggle Data Science Bowl released data from 700 patients compiled by the National Institutes of Health and Children's National Medical Center, and was at the time, an order of magnitude larger than any cardiac MRI dataset previously described. Patients were recruited from the United States and scans were performed in Washington DC area. While demographic splits from the dataset are not available, the original data was sourced from multiple hospital systems, across a range of age groups, containing both normal and diseased hearts. The competition closed on the 14$^{th}$ of March 2016, but data from 697 cases remain publicly available in the DICOM format.[39] 2CH, 4CH, and SAX cine sequences were available for use, along with expert annotations for left ventricular end systolic and end diastolic volumes. The entirety of the available dataset was used for external validation as is without any quality control.

## Neural Network Architectures:

We tested vision encoder architectures including 3D residual convolutional networks and video vision transformers. We settled on using an implementation of a multi-scale vision transformer (mViT) with 36.3 million trainable parameters as our video encoder after experiments showing superior generalization and embedding quality.[26] Vision transformers have recently emerged as a performant alternative to convolutional neural networks, especially in the setting of large scale self-supervised pre-training.[76,77] Vision transformers retain the skip connections seen in traditional convolutional networks, but are also able to attend to local and global features of an image in earlier stages.[78] The mViT architecture is a vision transformer designed specifically for video data that foregoes the successive layers of convolutional operations seen in typical convolutional neural networks, for a single convolutional layer to divide the input video into a linear series of overlapping cubes. These linear elements are processed by 16 layers of stacked transformer modules, allowing the network to effectively attend to distant input features. Specific to the mViT architecture is a sequential series of pooling and scaling operations, that effectively enable the network to attend to simple visual features at high resolution in early layers, followed complex high-dimensional relationships at a coarser resolution in deeper layers. As a result, compared to other extensions of 2D-image transformers to the video domain, mViT by design has a stronger temporal inductive bias. While more computationally expensive than comparable convolutional networks, mViT is more efficient than comparable vision transformers, requiring remarkably less pre-training data to achieve state-of-the art results on typical action recognition datasets. Finally, compared to traditional convolutional neural networks, mViT has shown superior performance on large video action recognition datasets despite fewer trainable parameters.[26]

We elected to use a pre-trained BERT (Bidirectional encoder representations from transformers) model for our text encoder.[27] Unlike other language models that have come before it, BERT is trained using a 'bidirectional' approach, wherein the model is trained to learn the structure and context of human language by attending to sentences in both the left-to-right and right-to-left direction. Specific details of the pretraining methods for BERT are detailed in the original paper.[27] We use a 12-layer variant of BERT-base, with 12 attention heads, and a hidden dimension of size 786, with a total of 110 million trainable parameters. We tested a combination of different pre-trained weights including those from the original publication, weights finetuned on the MIMIC dataset, and weights from a model trained on biomedical abstracts from PubMed with a custom vocabulary of 30522 tokens.[79,80] (Supplementary Fig. 3).

### Pre-training framework

We build on previous attempts at learning visual representations using naturally occurring pairing of two-dimensional medical imaging and textual data, extending these concepts to the spatiotemporal video-like nature of cardiac MRI scans.[14–17,19,20] Two parallel encoders were trained – one for processing the MRI cine-sequences and the other for processing the subsampled text from paired radiology reports. Self-supervised transformer networks in particular, have shown superior performance on downstream tasks compared to traditional supervised techniques.[76,81,82] We used an implementation of a Multi-scale Vision Transformer (mViT) with Kinetics-400 initialized weights for the vision encoder, and a pretrained BERT model for the text report encoder. Specifically, we utilized weights from BERT pre-trained on abstracts of biomedical publications on PubMed with a custom vocabulary.[79] Data from 8513 patients (9427 scans and paired reports) were used for training, and a separate set from 4194 patients (4646 scans and paired reports) were used for validation.

We employ randomized sequential data augmentation schemes (AugMix) to stochastically sample and layer a series of chained transformations including but not limited to resizing, solarization, shear, translate and random rotation of videos in the spatial dimensions, all while preserving the same augmentations along the temporal dimension for temporal consistency.[83] Uniform temporal subsampling greatly improved downstream performance and generalizability. We augment the radiology reports by randomly sampling five sentences from the entire report for each scan per training step. The output of each encoder is passed through a one-layer linear projection head to yield a pair of 512-dimensional embeddings. These low-dimensional 512-D embeddings are a compressed numeric representation of the information contained within the input MRI scan and paired text report.

Prior work has also shown the importance of large batch sizes for effective contrastive representation learning.[81] To study this we pretrained models with a batch size of 16, 32, and 128 video-text pairs. For the UK BioBank LVEF prediction task, we found a that finetuning from the larger batch-size pretrained models led to improved downstream results (Supplementary Fig. 2). While computational budgets did not allow for an extensive hyperparameter search with the larger batch sizes, we note that the downstream benefits did not appear to be clinically significant for this specific task. Nonetheless, this remains an area for additional future exploration.

Vision-only self supervised methods would be challenging to incorporate where scans from multiple visually distinct view-planes exist for the same patient. We focus our efforts on text-to-video approaches given the success with text supervised visual representation learning across radiology and action recognition.[14,16,84,85] We considered approaches such as CLIP, however these are limited by a short context length suitable for captions rather than the larger, mostly unstructured paragraphs that are typical of cardiac MRI reports.[85] Similar to the work by Yuhao *et. al,* we elected to use an asymmetric bidirectional implementation of the InfoNCE loss to maximize mutual information between each MRI video : text report pair.[16,22] The contrastive losses used are essentially log-loss of a N-way classifier to predict the correct pair of MRI scan and report (where N = batch size). The first loss function is a video-to-text contrastive loss for the $i^{th}$ pair, where $\boldsymbol{v_i}$ represents a video embedding, and $\boldsymbol{u_i}$ represents a text embedding of the $i^{th}$ video-text pair. $\boldsymbol{N}$ here represents the number of video-text pairs in a total batch being evaluated.

Eq (1)

$$l_i^{(v\rightarrow u)} = -\log \frac{\exp(\langle \mathbf{v_i}, \mathbf{u_i} \rangle / \tau)}{\sum_{k=1}^{N} \exp(\langle \mathbf{v_i}, \mathbf{u_k} \rangle / \tau)}$$

The second loss function is a similarly structured text-to-video contrastive loss. The tunable temperature parameter ($\boldsymbol{\tau}$) controls the strength of penalties on hard negative pairs sampled during training.[86]

The final loss was defined as a weighted combination of the two losses averaged over all positive video – text pairs in each batch of data. The scalar weight is given by $\boldsymbol{\lambda}$.

Eq (2)

$$\mathcal{L} = \frac{1}{N} \sum_{i=1}^{N} \left( \lambda l_i^{(v\rightarrow u)} + (1 - \lambda) l_i^{(u\rightarrow v)} \right)$$

We additionally implemented a 'flooding' regularization technique to prevent the training loss ($\mathcal{L}$) to approach zero.[87] We set the flood level (scalar value given by $\boldsymbol{b}$) to a training loss of 0.05 to allow for better generalization. The final loss ($\tilde{\mathcal{L}}$)is thus given by:

Eq (3)

$$\tilde{\mathcal{L}} = |\mathcal{L} - b| + b$$

The specific pre-trained weights and vocabulary used for initializing the text encoder, batch size, augmentation scheme, InfoNCE temperature parameter, and flood regularization were critical for model convergence.[88] The final model was pretrained with a batch size of 32 per GPU, for 600 epochs. The first six layers of the BERT text-encoder was frozen, and the entire network was trained with a learning rate of 4.8e-5 using the AdamW optimizer with weight decay set to 1e-6, eps set to 1e-8. We decayed the learning rate by a factor of 0.1 at 300 epochs. Checkpoints were saved every 10 epochs during the pre-training process, and last checkpoint was used for finetuning on downstream clinical tasks. The total time taken for pre-training was 13 days and 14 hours (4 x 80GB Nvidia A100 GPUs). The ability of the vision transformer encoder to cluster different disease conditions without any additional explicit supervised training was visualized using the UMAP algorithm initialized using default values.[89]

## Multi-instance self-attention and downstream evaluation:

A gated multiview self-attention network is trained to assign an attention value ($\mathbf{a_k}$) to each MRI view embedding produced by the main vision encoder.[13,31] For each embedding within a bag of $\boldsymbol{K}$ embeddings, high score after softmax activation (near 1) indicates that a particular MRI view plane is highly informative for the downstream diagnostic task. Conversely, a low score (near 0) indicates that the MRI view plane has little to no diagnostic value. For classification tasks, each input embedding is additionally passed through a LayerNorm function prior to a forward pass into the self-attention blocks (Supplementary Fig. 6).[90] ($\boldsymbol{w^T}$, weight parameters; $\boldsymbol{V}$, weight parameters; $\boldsymbol{h_i}$, low dimensional embeddings; $\odot$, element-wise product; tanh, tanH activation function; sigm, sigmoid activation function; N, total number of MRI view embeddings for a particular study).

Eq (4)

$$a_k = \frac{\exp\{\mathbf{w}^\top(\tanh(\mathbf{V}\mathbf{h}_k^\top) \odot \mathrm{sigm}(\mathbf{U}\mathbf{h}_k^\top))\}}{\sum_{j=1}^{K} \exp\left\{\mathbf{w}^\top\left(\tanh\left(\mathbf{V}\mathbf{h}_j^\top\right) \odot \mathrm{sigm}\left(\mathbf{U}\mathbf{h}_j^\top\right)\right)\right\}}$$

We make use of an attention pooling mechanism to average the embeddings from all MRI views weighted by their predicted attention scores, to return a single 512-dimensional embedding. This embedding can be treated as a 'feature representation' of the entire MRI study for a specific downstream task of interest. For each downstream classification task of interest, we use a binary classification head with a sigmoid activation function, as disease labels are usually not mutually exclusive in the setting of cardiovascular disorders. For downstream tasks that involve regression of a numeric variable, we replace the binary classification head with a single output neuron with a linear activation function.

LVEF regression task

We examine two modes of training for LVEF% prediction: 1. 'Finetuning' where the last linear layer of the vision encoder and classifier head are trainable, and 2. 'Transfer learning' where the vision encoder, linear layer, and classifier heads are all trainable. 'Finetuning' allows for some degree of flexibility in the way embeddings are generated but keeps the vision encoder frozen to make use of the learned representations. With the system set to 'Transfer learning', the network begins from the learned representations, however since the entire network is unfrozen it is possible to 'overwrite' these parameters with each new update of the training process. For these experiments we initialize the vision encoder with the contrastive pretrained (ours) or Kinetics-400 (baseline), and onto which we attach the regression head as described above.

We finetune our pre-trained checkpoints with 32-bit precision using the AdamW optimizer, with a learning rate set to 1e-4, and default value of 0.01 for weight decay. We explored different augmentation schemes and achieved superior validation performance with AugMix on restricted hyperparameter sweeps with 10% of the training data.[83] For all experiments involving finetuning with subsets of available data, we use a manual seed value for random subsampling to ensure reproducibility of results. We make use of all available 4CH, 2CH, 3CH, and a random subsample of 50% of SAX views per study with no manual screening for quality control. We elect to train our regression models with a Huber Loss function, and we use mean squared errors and mean absolute errors as performance metrics[91]. We additionally calculate the area under receiver operator curve (AUROC) for diagnosing heart failure based on a LVEF cut-off of 40%. We train models for a maximum of 100,000 steps on GPUs with at least 16GB of VRAM each. For experiments described in Fig 3a and 3d, configuration files were generated for each experimental setup and were trained in parallel across numerous GPUs on Stanford Sherlock.

Disease classification task

We define both a 'finetuning' and 'transfer learning' as above, and use the same network architecture initialized with Kinetics-400 weights as our baseline. We finetune our pretrained checkpoints with the same overall settings as described above for the regression tasks, except using a weight decay value of 5e-4 and the addition of a LayerNorm function for the embeddings prior to a forward pass through the multi-instance self-attention modules to aid with convergence. We empirically use AugMix for our data augmentation strategy given the successes noted above. We make use of all available 4CH, 2CH, 3CH, and SAX views per study with no quality control or screening. We utilize a binary cross-entropy loss function with a sigmoid activation weighted by a scalar multiplier equal to the proportion of positive vs negative classes for each disease (calculated using the internal training set prevalences). We use the AUROC as a performance metric given the considerable class imbalance for positive and negative classes.[92] For each disease label of interest, we train models for 24 epochs, on GPUs with at least 24GB of VRAM. For experiments described in Figs. 4a and 4b configuration files were generated for each experimental setup and were trained in parallel across numerous GPUs on Stanford Sherlock. External test data were evaluated on the Penn CBICA cluster on a single Nvidia A40 GPU with 40GB VRAM and the PARCC Betty cluster on a single Nvidia Blackwell B200 GPU with 180GB VRAM. In addition to the losses and metrics, we store predicted probabilities and relative self-attention scores for each view for downstream processing and statistical analyses.

Statistical analyses:

We use the torchmetrics (v 1.0.1) package to calculate mean squared errors (mse), mean absolute errors (mae) for regression tasks, and AUROC values for classification tasks within the training and validation loops. We additionally manually calculate AUROCs as empirical curves in the sensitivity and specificity space, computed from predicted probabilities generated by our models.[93] To compare the performance of finetuned classifier models (ie Contrastive pre-trained vs Baseline), we calculated non-parametric confidence intervals on the AUROC using DeLong's method (paired),[94] following which p-values were computed for the mean difference between AUROC curves. Additional analyses are performed to calculate accuracy for each diagnostic label at different thresholds (optimizing for Youden's statistic, a sensitivity of 0.90 or a specificity of 0.90). Differences between predicted LVEF% values and ground-truth were assessed using Bland & Altman plots. Statistical analyses were performed, and graphs were plotted using R (v4.1.0); major packages used include pROC (v 1.17.0), ggplot2 (3.3.5), blandr (0.5.1). The online test-set leaderboard webapp was created using shiny (1.8.1).

Attention visualizations:

For every input scan, we output the raw self-attention tensors from each head of each layer of the MRI vision encoder during evaluation, and process them to yield 65 separate attention heatmaps. As described earlier, the spatiotemporal resolution reduces with each successive stage in the MVIT architecture; the self-attention tensors reduce from an initial spatiotemporal resolution of [8 x 56 x 56] at the first layer, to [8 x 7 x 7] at the last few layers. We keep only the attention values from the output patches for the purposes of visualization, and spatiotemporally interpolate these tensors back to a size of [16 x 224 x 224] via nearest neighbor resampling. These arrays are exported to mp4 files using imageio and the ffmpeg library (Supplementary Fig. 17 and 18). Aside from the self-attention heatmaps for each input video, we also compute the raw self-attention values from the multi-instance classifier head for relevant downstream tasks. After each scan is passed through the vision encoder, the resultant embedding is assigned a leaned raw self-attention score within the multi-instance self-attention modules. We calculate the relative differences in self-attention scores across different view planes for each disease label. These relative self-attention values are visualized as 2d heatmaps as shown in Fig 4c. The multi-instance classifier head self-attention scores show that the network learns to differentially prioritize view-planes for different clinical tasks.

**Data availability:**

Cardiac MRI Data from Stanford Medicine, UCSF, and Medstar were secured via agreements with Bunkerhill Health. Applications for data access can be found at https://www.bunkerhillhealth.com. Cardiac MRI data from the University of Pennsylvania are available to researchers under appropriate data use agreements. Please contact Rohan Shad, MD for additional information. Applications for data access from the UK Bio-Bank can be found at https://www.ukbiobank.ac.uk/enable-your-research/apply-for-access. The ACDC dataset is publicly available at https://www.creatis.insa-lyon.fr/Challenge/acdc/databases.html. The Kaggle MRI dataset is publicly available at https://www.kaggle.com/competitions/second-annual-data-science-bowl/data. Leaderboard for classification and regression test datasets are available at https://rohanshad.shinyapps.io/cmr_leaderboard/. We are happy to accept model weights as submissions to evaluate on our existing datasets, please contact Rohan Shad, MD for additional information.

**Code Availability:**

Repository for this project containing model checkpoints, raw table and figure data, and additional scripts to preprocess and analyze data are available at https://github.com/rohanshad/cmr_transformer. Final contrastive pretrained checkpoint for our models are available for academic use: https://huggingface.co/rohanshad/cmr_c0.1. Please contact the corresponding and first authors for commercial licenses. Additional scripts and preprocessing code for handling Cardiac MRI data are available free of cost at https://github.com/rohanshad/cmr_toolkit.

**Acknowledgements:**

Some of the computing for this project was performed on the Stanford Sherlock high performance computing cluster and the Stanford Nero Google Cloud Platform. We would like to thank Stanford University and the Stanford Research Computing Center for providing computational resources and support. Computing for external validation on the UPenn data was in performed on the Penn CUBIC, and PARCC Betty clusters. Part of this research has been conducted using the UK Biobank Resource under Application Number 71226. We would like to thank Yuhao Zhang (Stanford) for helpful discussions.

R.S was supported in part by University of Pennsylvania CBCIA Prodev Award for CUBIC Cluster usage, and the American Heart Association Postdoctoral Fellowship Award #834986. W.H is supported in part by NIH R01HL157235. R.S. and W.H. are supported, in part, by a CAROL Act supplement from the NHLBI (3R01HL157235-04S1). W.R.W. is supported in part by R01 HL137984, R01 HL169378 and P41EB029460. We would also like to thank Penn Radiology RADAR Service and the Penn Medicine Academic Computing Services. The funders had no role in study design, data collection, analysis, decision to publish or preparation of the manuscript.

**Author Contributions:**

R.S and W.H conceived of the methods, designed the study and analyses, interpreted the results and wrote the manuscript. R.S built the codebase and conducted all the experiments. C.Z, D.K, J.C, M.M, R.F, M.L assisted with the implementation of codebase and analysis of results. N.K, D.E, K.K, R.W.F, W.R.W, J.M, M.A.A, and R.F were involved with data collection. All authors reviewed the manuscript. W.H, W.R.W, R.S, M.A.A acquired funding to support the work. E.A.A, A.D.F, V.A.F, K.K, J.M, R.W.F, C.L, provided clinical supervision and assisted with the design of evaluation tasks. W.H supervised the overall research project.

**Competing Interests Statement:**

N.K and D.E are major shareholders of Bunkerhill Health. The remaining authors declare no competing interests.

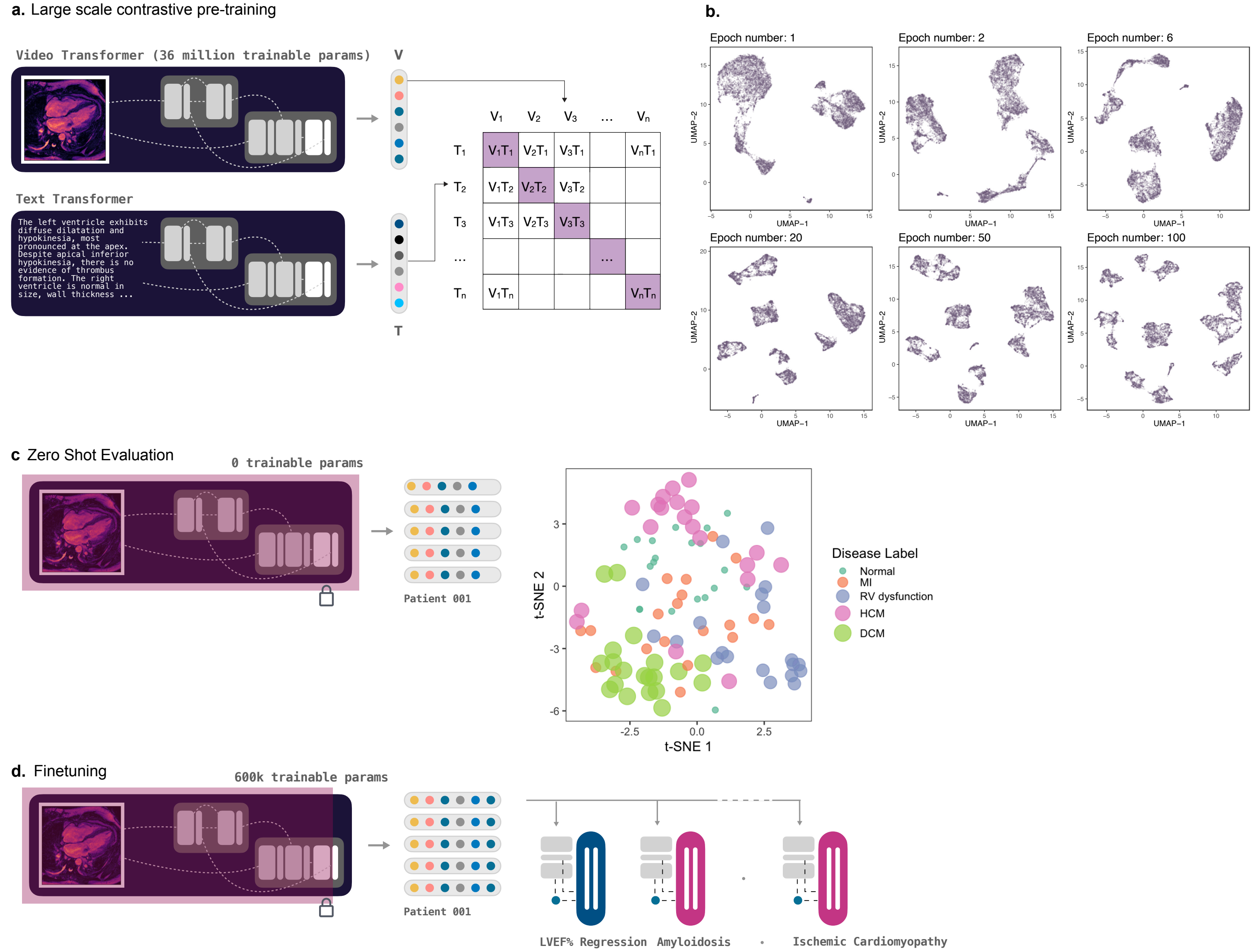

**Fig 1 Project overview. a.** Large scale contrastive pre-training. CMR cine-sequences in the form of video data are fed to a video transformer network, and the paired text reports are fed into a parallel text transformer network. The embeddings produced by each (V – video, T – text) represent compressed representations of the complex inputs to each network. The networks are trained to maximize agreement between true pairs of video and text that arise from the same CMR scan (ie. the diagonal of the similarity matrix highlighted: $V_1T_1$, $V_2T_2$ ... $V_nT_n$), and minimize the agreement between samples that come from different scans. In this process, we see (**b**), that as training progresses more complex local and global structural features emerge in the video embeddings (V). The scans initially viewed as an indistinct amalgam in high dimensional space at the start of training (epoch #1) begin to develop separations and localized clusters that are more distinct by epoch #100. **c**. Evaluating the ability of the network on external datasets in separating diseases or probing the UK BioBank as described in later sections is achieved by freezing all the parameters of the video encoder (at this point the text network is discarded). Embeddings produced from different views can then be plotted in 2-dimensions via standard dimension reduction algorithms. Finally, as shown in **d**., leaving the last layer unfrozen (about 600,000 trainable parameters), we showcase data-efficient finetuning towards specific clinical tasks of interest. Of note, the embeddings are fed into a secondary network designed to aggregate information from different views into a single prediction, listed above are a few: Regression of left ventricular ejection fraction (LVEF%), and diagnosis of Amyloidosis or Ischemic Cardiomyopathy. A single vision model can thus be applied to a wide range of different tasks.

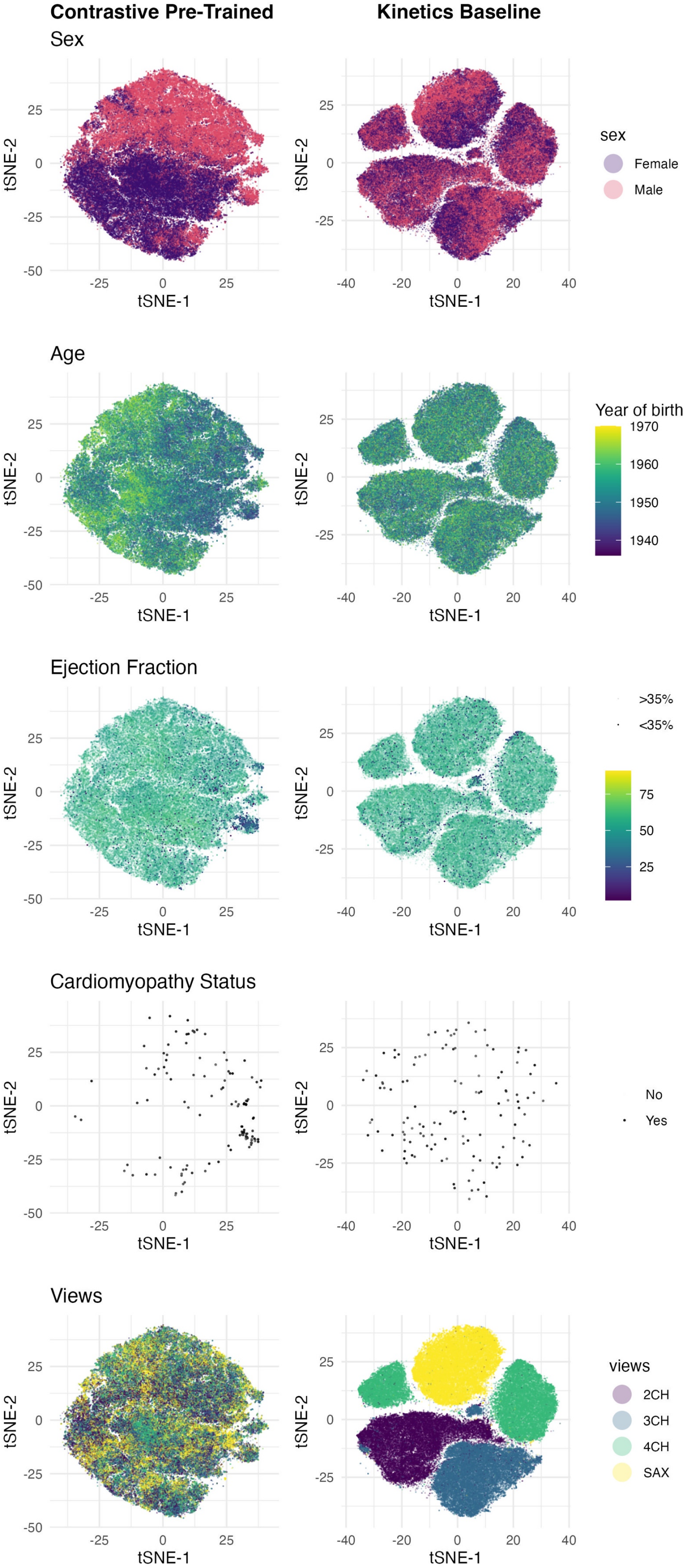

***Fig 2 Contrastive pre-training enables learning of demographic and pathophysiological representations.*** *2D t-SNE plots of low-dimensional embeddings generated from a forward pass on 34490 unique scans from the UK BioBank (n=31693 unique individuals) using a Kinetics-400 pre-trained checkpoint (baseline; right column) compared to a Contrastive pre-trained (ours; left column). The color labels help with understanding the basis of clustering. While at first glance it may appear that the Kinetics baseline model produces a set of five readily separable clusters vs the more homogenous appearing plot for the contrastive pre-trained model, the basis of the baseline Kinetics initialized clusters are simply the various MRI view planes. On further exploration we find that the Kinetics-400 generated embeddings fail to capture the required information to separate low ejection fraction states, cardiomyopathy, gender, or age (lack of separation of color for each variable). This is different from the contrastive pre-trained generated embeddings that show clear demarcations by physiologically relevant characteristics allowing for zero-shot separation of low-ejection fraction states, cardiomyopathy, gender, and age. Color legends for each subplot shown on the right of each t-SNE figure. View-invariance is a built-in feature of the contrastive pre-training process, evidenced by the characteristic absence of clustering by view-plane.*

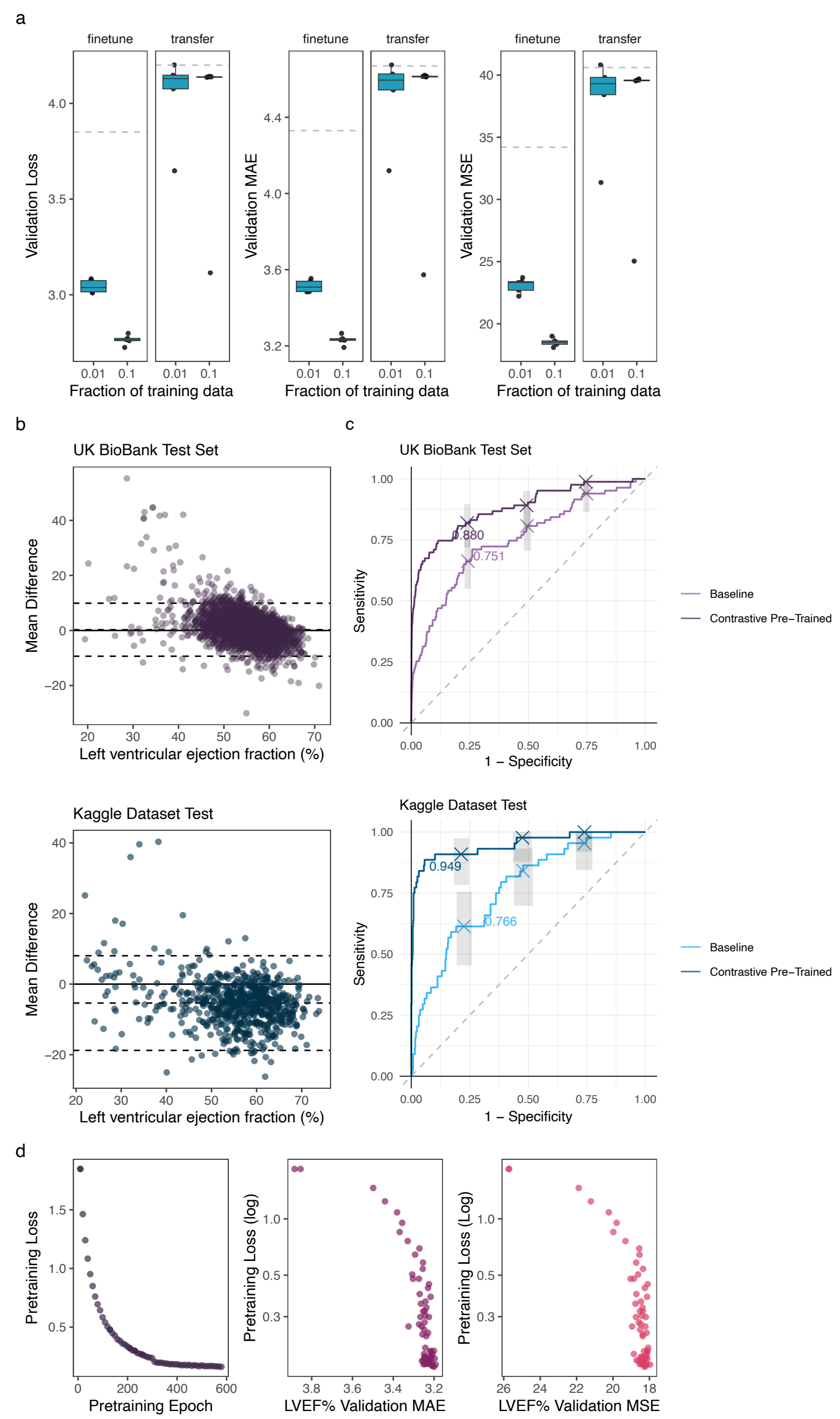

***Fig 3 Clinical grade LVEF% regression using contrastive pre-trained vision encoder. a.*** *Dashed horizontal line represents Kinetics-400 baseline performance for the problem of LVEF% regression with 100% of the UK BioBank training data (for finetuning and transfer learning modes). Each point represents validation-set results from a randomly initialized subset (n=5 independent replicates) of either 1% or 10% of the available training data (x-axis). Superior performance (lower is better) is seen with finetuning, whereas unfreezing all mViT layers (transfer learning) is detrimental to performance. Finetuning with just 1% of available data yields superior validation results compared to Kinetics-400 baseline. Box limits represent IQR, whiskers 1.5x IQR, horizontal marker is the median.* ***b****. Bland-Altman plots for test-set results on the UK BioBank, and an additional external test-set using the Kaggle Dataset. MAE on UK BioBank Test Set is 3.344 (sd 3.615), with BA limits of agreement -9.91% to +9.61%. Kaggle Test results: MAE of 6.880 (sd 5.309), with BA limits of agreement -18.7% to + 8.03%.* ***c.*** *Test-set performance of Kinetics-400 baseline performance against contrastive pre-trained models with 100% of training data available for HFrEF (LVEF < 40%) diagnosis based on predicted LVEF%. AUC on the UK BioBank test set is 0.880 (95%CI: 0.835-0.925; n=4259 scans), and 0.949 (95% CI 0.907-0.990; n=679 scans) on Kaggle dataset. Baseline performance on the UK BioBank test set was 0.751 (95% CI: 0.692-0.811; Z=3.93; $p=8.376E^{-5}$) and for the Kaggle dataset 0.766 (95% CI: 0.697-0.836; Z=4.08; $p=4.325E^{-5}$). Grey shaded regions represent 95% confidence intervals at different classifier thresholds. Statistical testing perfomed as two-sided paired DeLong's tests without multiple comparison adjustments.* ***d.*** *Panels illustrate relationship of downstream performance with quality of contrastive pre-training. Each point represents a pre-trained model checkpoint saved every 10 epochs of pre-training. Panels from left to right: graph showing non-monotonic decrease in loss as pre-training continues; progressive improvement in downstream performance on LVEF% regression task validation MAE; similar improvement in validation MSE as a function of decreasing pre-training loss. For this experiment fine-tuning is performed using 10% of available UK BioBank data.*

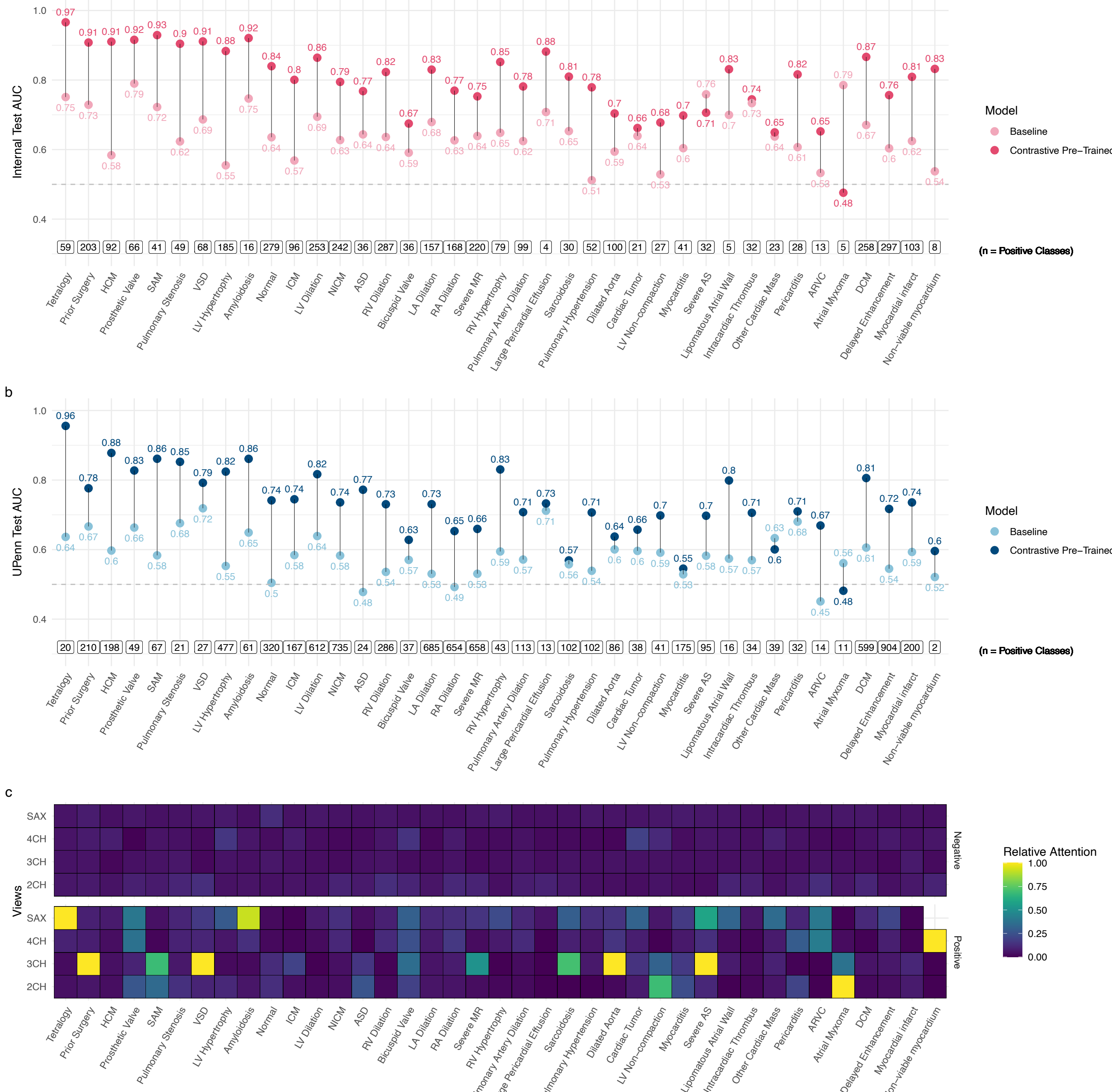

***Fig 4 Data efficient disease diagnosis from cine-sequences alone**. **a**. Test-set AUCs on the internal dataset for contrastive pre-trained (magenta) and baseline Kinetics-400 baseline (light pink) checkpoints for models fine-tuned for the binary classification problem of disease diagnosis. Disease labels are placed along the x-axis, with test-AUC on the y-axis. The dashed grey line marks AUC of 0.5 - the performance expected from a random classifier. Solid grey connecting lines mark the performance difference between contrastive pre-trained and baseline models Remarkable classification performance is observed for conditions such as Hypertrophic cardiomyopathy, Amyloidosis, Systolic Anterior motion of Mitral valve, and Dilated Cardiomyopathy. Whereas models somewhat struggle for conditions that typically require gadolinium contrast or additional scanning sequences for diagnosis such as differentiating cardiac masses from cardiac thrombus**.** The number of patients with disease for each disease label are listed along the x-axis (n = Positive Classes), in black text within grey outlined box **b.** Test-set results of the contrastive pre-trained (navy-blue) and Kinetics-400 baseline (light blue) models on the UPenn dataset with no additional fine-tuning**. c.** Relative self-attention values calculated by the multi-instance self-attention classifier heads. Heatmaps generated for different views (2CH, 3CH, 4CH, SAX) for each disease condition, plotted as a function of the presence or absence of the disease in question. Subjectively we find that the networks offer more attention to certain views over others depending on the disease condition to make accurate predictions. This is additionally consistent with internal testing that showed performance degradation when certain slices or views are hidden from the networks.*

**References:**

1. Salerno, M. *et al.* Recent Advances in Cardiovascular Magnetic Resonance: Techniques and Applications. *Circ Cardiovasc Imaging* **10**, e003951 (2017).
2. Leiner, T. *et al.* SCMR Position Paper (2020) on clinical indications for cardiovascular magnetic resonance. *J Cardiovasc Magn Reson* **22**, 76 (2020).
3. Poplin, R. *et al.* Prediction of cardiovascular risk factors from retinal fundus photographs via deep learning. *Nature Biomedical Engineering* **2**, 158–164 (2018).
4. Ouyang, D. *et al.* Video-based AI for beat-to-beat assessment of cardiac function. *Nature* 1–5 (2020) doi:10.1038/s41586-020-2145-8.
5. Popescu, D. M. *et al.* Arrhythmic sudden death survival prediction using deep learning analysis of scarring in the heart. *Nat Cardiovasc Res* https://doi.org/10.1038/s44161-022-00041-9 (2022) doi:10.1038/s44161-022-00041-9.
6. Hannun, A. Y. *et al.* Cardiologist-level arrhythmia detection and classification in ambulatory electrocardiograms using a deep neural network. *Nature Medicine* **25**, 65–69 (2019).
7. Shad, R. *et al.* Predicting post-operative right ventricular failure using video-based deep learning. *Nat Commun* **12**, 5192 (2021).
8. Beede, E. *et al.* A Human-Centered Evaluation of a Deep Learning System Deployed in Clinics for the Detection of Diabetic Retinopathy. in *Proceedings of the 2020 CHI Conference on Human Factors in Computing Systems* 1–12 (ACM, Honolulu HI USA, 2020). doi:10.1145/3313831.3376718.
9. Geirhos, R. *et al.* Shortcut learning in deep neural networks. *Nature Machine Intelligence* **2**, 665–673 (2020).
10. Jafari, M. *et al.* Automated diagnosis of cardiovascular diseases from cardiac magnetic resonance imaging using deep learning models: A review. *Computers in Biology and Medicine* **160**, 106998 (2023).
11. Carr, J. C. *et al.* Cine MR Angiography of the Heart with Segmented True Fast Imaging with Steady-State Precession. *Radiology* **219**, 828–834 (2001).
12. Bieri, O. & Scheffler, K. Fundamentals of balanced steady state free precession MRI: Fundamentals of Balanced SSFP MRI. *J. Magn. Reson. Imaging* **38**, 2–11 (2013).
13. Lu, M. Y. *et al.* AI-based pathology predicts origins for cancers of unknown primary. *Nature* **594**, 106–110 (2021).
14. Tiu, E. *et al.* Expert-level detection of pathologies from unannotated chest X-ray images via self-supervised learning. *Nat. Biomed. Eng* https://doi.org/10.1038/s41551-022-00936-9 (2022) doi:10.1038/s41551-022-00936-9.
15. Chen, R. J. *et al.* Towards a general-purpose foundation model for computational pathology. *Nat Med* **30**, 850–862 (2024).
16. Zhang, Y., Jiang, H., Miura, Y., Manning, C. D. & Langlotz, C. P. Contrastive Learning of Medical Visual Representations from Paired Images and Text. *arXiv:2010.00747 [cs]* http://arxiv.org/abs/2010.00747 (2020).
17. Lu, M. Y. *et al.* A visual-language foundation model for computational pathology. *Nat Med* **30**, 863–874 (2024).
18. Taleb, A. *et al.* 3D Self-Supervised Methods for Medical Imaging. *arXiv:2006.03829v3 [cs.CV]* 19.
19. Azizi, S. *et al.* Big Self-Supervised Models Advance Medical Image Classification. Preprint at https://doi.org/10.48550/arXiv.2101.05224 (2021).
20. Krishnan, R., Rajpurkar, P. & Topol, E. J. Self-supervised learning in medicine and healthcare. *Nat. Biomed. Eng* **6**, 1346–1352 (2022).
21. Zhou, Y. *et al.* A foundation model for generalizable disease detection from retinal images. *Nature* **622**, 156–163 (2023).
22. Oord, A. van den, Li, Y. & Vinyals, O. Representation Learning with Contrastive Predictive Coding. *arXiv:1807.03748 [cs, stat]* http://arxiv.org/abs/1807.03748 (2019).
23. Vukadinovic, M. *et al.* EchoPrime: A Multi-Video View-Informed Vision-Language Model for Comprehensive Echocardiography Interpretation. Preprint at https://doi.org/10.48550/arXiv.2410.09704 (2024).
24. Blankemeier, L. *et al.* Merlin: A Vision Language Foundation Model for 3D Computed Tomography. Preprint at https://doi.org/10.48550/arXiv.2406.06512 (2024).
25. Beeche, C. *et al.* A Pan-Organ Vision-Language Model for Generalizable 3D CT Representations. Preprint at https://doi.org/10.1101/2025.07.03.25330654 (2025).
26. Fan, H., Xiong, B. & Mangalam, K. Multiscale Vision Transformers. *arXiv:2104.11227 [cs.CV]* 18 (2021).

27. Devlin, J., Chang, M.-W., Lee, K. & Toutanova, K. BERT: Pre-training of Deep Bidirectional Transformers for Language Understanding. *arXiv:1810.04805 [cs]* http://arxiv.org/abs/1810.04805 (2019).
28. Bernard, O. *et al.* Deep Learning Techniques for Automatic MRI Cardiac Multi-Structures Segmentation and Diagnosis: Is the Problem Solved? *IEEE Trans. Med. Imaging* **37**, 2514–2525 (2018).
29. Carreira, J., Noland, E., Banki-Horvath, A., Hillier, C. & Zisserman, A. A Short Note about Kinetics-600. *arXiv:1808.01340 [cs]* http://arxiv.org/abs/1808.01340 (2018).
30. Maaten, L. van der & Hinton, G. Visualizing data using t-SNE. *Journal of machine learning research* **9**, 2579–2605 (2008).
31. Ilse, M., Tomczak, J. M. & Welling, M. Attention-based Deep Multiple Instance Learning. *arXiv:1802.04712 [cs, stat]* http://arxiv.org/abs/1802.04712 (2018).
32. Gheorghiță, B. A. *et al.* Improving robustness of automatic cardiac function quantification from cine magnetic resonance imaging using synthetic image data. *Sci Rep* **12**, 2391 (2022).
33. Liao, F., Chen, X., Hu, X. & Song, S. Estimation of the volume of the left ventricle from MRI images using deep neural networks. *IEEE Trans. Cybern.* **49**, 495–504 (2019).
34. Bai, W. *et al.* Automated cardiovascular magnetic resonance image analysis with fully convolutional networks. *J Cardiovasc Magn Reson* **20**, 65 (2018).
35. Fu, Y. *et al.* A versatile foundation model for cine cardiac magnetic resonance image analysis tasks. Preprint at https://doi.org/10.48550/arXiv.2506.00679 (2025).
36. Zhang, Y. *et al.* Towards Cardiac MRI Foundation Models: Comprehensive Visual-Tabular Representations for Whole-Heart Assessment and Beyond. *Medical Image Analysis* **106**, 103756 (2025).
37. Bhuva, A. N. *et al.* A Multicenter, Scan-Rescan, Human and Machine Learning CMR Study to Test Generalizability and Precision in Imaging Biomarker Analysis. *Circ: Cardiovascular Imaging* **12**, e009214 (2019).
38. Suinesiaputra, A. *et al.* Quantification of LV function and mass by cardiovascular magnetic resonance: multi-center variability and consensus contours. *J Cardiovasc Magn Reson* **17**, 63 (2015).
39. Booz Allen Hamilton. Kaggle Second Annual Data Science Bowl. https://www.kaggle.com/c/second-annual-data-science-bowl/ (2015).
40. Nikolaidou, C. & Karamitsos, T. Should everyone have an MRI in heart failure? *Cardiovasc Diagn Ther* **10**, 549–553 (2020).
41. WRITING COMMITTEE MEMBERS *et al.* ACCF/ACR/AHA/NASCI/SCMR 2010 Expert Consensus Document on Cardiovascular Magnetic Resonance: A Report of the American College of Cardiology Foundation Task Force on Expert Consensus Documents. *Circulation* **121**, 2462–2508 (2010).
42. Mancio, J. *et al.* Machine learning phenotyping of scarred myocardium from cine in hypertrophic cardiomyopathy. *European Heart Journal - Cardiovascular Imaging* **23**, 532–542 (2022).
43. Zhang, N. *et al.* Deep Learning for Diagnosis of Chronic Myocardial Infarction on Nonenhanced Cardiac Cine MRI. *Radiology* **291**, 606–617 (2019).
44. Weinsaft, J. W. *et al.* Contrast-Enhanced Anatomic Imaging as Compared to Contrast-Enhanced Tissue Characterization for Detection of Left Ventricular Thrombus. *JACC: Cardiovascular Imaging* **2**, 969–979 (2009).
45. O'Donnell, D. H. *et al.* Cardiac Tumors: Optimal Cardiac MR Sequences and Spectrum of Imaging Appearances. *American Journal of Roentgenology* **193**, 377–387 (2009).
46. Wang, T. K. M. *et al.* Cardiac Magnetic Resonance Imaging Techniques and Applications for Pericardial Diseases. *Circ: Cardiovascular Imaging* **15**, (2022).
47. Eichhorn, C. *et al.* Multiparametric Cardiovascular Magnetic Resonance Approach in Diagnosing, Monitoring, and Prognostication of Myocarditis. *JACC: Cardiovascular Imaging* **15**, 1325–1338 (2022).
48. Bosman, L. P. *et al.* Diagnosing arrhythmogenic right ventricular cardiomyopathy by 2010 Task Force Criteria: clinical performance and simplified practical implementation. *EP Europace* **22**, 787–796 (2020).
49. Mathew, R. C., Löffler, A. I. & Salerno, M. Role of Cardiac Magnetic Resonance Imaging in Valvular Heart Disease: Diagnosis, Assessment, and Management. *Curr Cardiol Rep* **20**, 119 (2018).
50. Naidu, S. S. *et al.* Frequency and clinicoeconomic impact of delays to definitive diagnosis of obstructive hypertrophic cardiomyopathy in the United States. *Journal of Medical Economics* **26**, 682–690 (2023).
51. Witteles, R. M. *et al.* Screening for Transthyretin Amyloid Cardiomyopathy in Everyday Practice. *JACC: Heart Failure* https://doi.org/10.1016/j.jchf.2019.04.010 (2019) doi:10.1016/j.jchf.2019.04.010.
52. Maurer Mathew S. *et al.* Tafamidis Treatment for Patients with Transthyretin Amyloid Cardiomyopathy. *New England Journal of Medicine* **379**, 1007–1016 (2018).

53. Olivotto, I. *et al.* Mavacamten for treatment of symptomatic obstructive hypertrophic cardiomyopathy (EXPLORER-HCM): a randomised, double-blind, placebo-controlled, phase 3 trial. *The Lancet* **396**, 759–769 (2020).
54. Anthony, C. *et al.* Cardiovascular Magnetic Resonance for Rejection Surveillance After Cardiac Transplantation. *Circulation* **145**, 1811–1824 (2022).
55. Kaur, D. *et al.* Abstract 4124675: Deep Learning Screening of Cardiac MRIs Uncovers Undiagnosed Hypertrophic Cardiomyopathy in the UK BioBank. *Circulation* **150**, (2024).
56. Petersen, S. E. *et al.* UK Biobank's cardiovascular magnetic resonance protocol. *J Cardiovasc Magn Reson* **18**, 8 (2015).
57. Flynn, B. I. *et al.* Deep learning based phenotyping of medical images improves power for gene discovery of complex disease. *npj Digit. Med.* **6**, 155 (2023).
58. Pirruccello, J. P. *et al.* *Genetic Analysis of Right Heart Structure and Function in 40,000 People*. http://biorxiv.org/lookup/doi/10.1101/2021.02.05.429046 (2021) doi:10.1101/2021.02.05.429046.
59. Bonazzola, R. *et al.* Image-Derived Phenotype Extraction for Genetic Discovery via Unsupervised Deep Learning in CMR Images. in *Medical Image Computing and Computer Assisted Intervention – MICCAI 2021* (eds De Bruijne, M. et al.) vol. 12905 699–708 (Springer International Publishing, Cham, 2021).
60. Aung, N. *et al.* Genome-Wide Analysis of Left Ventricular Image-Derived Phenotypes Identifies Fourteen Loci Associated With Cardiac Morphogenesis and Heart Failure Development. *Circulation* **140**, 1318–1330 (2019).
61. Thanaj, M. *et al.* Genetic and environmental determinants of diastolic heart function. *Nat Cardiovasc Res* **1**, 361–371 (2022).
62. Aung, N. *et al.* Genome-wide association analysis reveals insights into the genetic architecture of right ventricular structure and function. *Nat Genet* **54**, 783–791 (2022).
63. Ho, C. Y. *et al.* Genotype and Lifetime Burden of Disease in Hypertrophic Cardiomyopathy: Insights From the Sarcomeric Human Cardiomyopathy Registry (SHaRe). *Circulation* **138**, 1387–1398 (2018).
64. Charron, P. *et al.* The Cardiomyopathy Registry of the EURObservational Research Programme of the European Society of Cardiology: baseline data and contemporary management of adult patients with cardiomyopathies. *European Heart Journal* **39**, 1784–1793 (2018).
65. Witschey, W. R. T. *et al.* Medical image analysis platform and associated methods. (2024).
66. BehnamGhader, P. *et al.* LLM2Vec: Large Language Models Are Secretly Powerful Text Encoders. Preprint at http://arxiv.org/abs/2404.05961 (2024).
67. Martini, N. *et al.* Deep learning to diagnose cardiac amyloidosis from cardiovascular magnetic resonance. *J Cardiovasc Magn Reson* **22**, 84 (2020).
68. Hinojar, R. *et al.* T1 Mapping in Discrimination of Hypertrophic Phenotypes: Hypertensive Heart Disease and Hypertrophic Cardiomyopathy: Findings From the International T1 Multicenter Cardiovascular Magnetic Resonance Study. *Circ: Cardiovascular Imaging* **8**, e003285 (2015).
69. Ommen, S. R. *et al.* 2024 AHA/ACC/AMSSM/HRS/PACES/SCMR Guideline for the Management of Hypertrophic Cardiomyopathy: A Report of the American Heart Association/American College of Cardiology Joint Committee on Clinical Practice Guidelines. *Circulation* CIR.0000000000001250 (2024) doi:10.1161/CIR.0000000000001250.
70. Paszke, A. *et al.* PyTorch: An Imperative Style, High-Performance Deep Learning Library. *arXiv:1912.01703 [cs, stat]* http://arxiv.org/abs/1912.01703 (2019).
71. Qi, P., Zhang, Y., Zhang, Y., Bolton, J. & Manning, C. D. Stanza: A Python Natural Language Processing Toolkit for Many Human Languages. *arXiv:2003.07082 [cs]* http://arxiv.org/abs/2003.07082 (2020).
72. Smit, A. *et al.* CheXbert: Combining Automatic Labelers and Expert Annotations for Accurate Radiology Report Labeling Using BERT. Preprint at http://arxiv.org/abs/2004.09167 (2020).
73. Zakka, C., Chaurasia, A., Shad, R. & Hiesinger, W. Almanac: Knowledge-Grounded Language Models for Clinical Medicine. Preprint at http://arxiv.org/abs/2303.01229 (2023).
74. Sellergren, A. *et al.* MedGemma Technical Report. Preprint at https://doi.org/10.48550/arXiv.2507.05201 (2025).
75. Petersen, S. E. *et al.* Imaging in population science: cardiovascular magnetic resonance in 100,000 participants of UK Biobank - rationale, challenges and approaches. *J Cardiovasc Magn Reson* **15**, 46 (2013).
76. Caron, M. *et al.* Emerging Properties in Self-Supervised Vision Transformers. *arXiv:2104.14294 [cs]* http://arxiv.org/abs/2104.14294 (2021).

77. Dosovitskiy, A. *et al.* An Image is Worth 16x16 Words: Transformers for Image Recognition at Scale. *arXiv:2010.11929 [cs]* http://arxiv.org/abs/2010.11929 (2020).
78. Raghu, M., Unterthiner, T., Kornblith, S., Zhang, C. & Dosovitskiy, A. Do Vision Transformers See Like Convolutional Neural Networks? Preprint at http://arxiv.org/abs/2108.08810 (2022).
79. Gu, Y. *et al.* Domain-Specific Language Model Pretraining for Biomedical Natural Language Processing. *ACM Trans. Comput. Healthcare* **3**, 1–23 (2022).
80. Alsentzer, E. *et al.* Publicly Available Clinical BERT Embeddings. in *Proceedings of the 2nd Clinical Natural Language Processing Workshop* 72–78 (Association for Computational Linguistics, Minneapolis, Minnesota, USA, 2019). doi:10.18653/v1/W19-1909.
81. Chen, T., Kornblith, S., Norouzi, M. & Hinton, G. A Simple Framework for Contrastive Learning of Visual Representations. Preprint at http://arxiv.org/abs/2002.05709 (2020).
82. He, K., Fan, H., Wu, Y., Xie, S. & Girshick, R. Momentum Contrast for Unsupervised Visual Representation Learning. Preprint at http://arxiv.org/abs/1911.05722 (2020).
83. Hendrycks, D. *et al.* AugMix: A Simple Data Processing Method to Improve Robustness and Uncertainty. Preprint at http://arxiv.org/abs/1912.02781 (2020).
84. Eyuboglu, S. *et al.* Multi-task weak supervision enables anatomically-resolved abnormality detection in whole-body FDG-PET/CT. *Nat Commun* **12**, 1880 (2021).
85. Radford, A. *et al.* Learning Transferable Visual Models From Natural Language Supervision. Preprint at http://arxiv.org/abs/2103.00020 (2021).
86. Wang, F. & Liu, H. Understanding the Behaviour of Contrastive Loss. *arXiv:2012.09740 [cs]* http://arxiv.org/abs/2012.09740 (2021).
87. Ishida, T., Yamane, I., Sakai, T., Niu, G. & Sugiyama, M. Do We Need Zero Training Loss After Achieving Zero Training Error? *arXiv:2002.08709 [cs, stat]* http://arxiv.org/abs/2002.08709 (2021).
88. Nakkiran, P. *et al.* Deep Double Descent: Where Bigger Models and More Data Hurt. *arXiv:1912.02292 [cs, stat]* http://arxiv.org/abs/1912.02292 (2019).
89. McInnes, L., Healy, J. & Melville, J. UMAP: Uniform Manifold Approximation and Projection for Dimension Reduction. http://arxiv.org/abs/1802.03426 (2018).
90. Brody, S., Alon, U. & Yahav, E. On the Expressivity Role of LayerNorm in Transformers' Attention. Preprint at http://arxiv.org/abs/2305.02582 (2023).
91. Huber, P. J. Robust Estimation of a Location Parameter. *Ann. Math. Statist.* **35**, 73–101 (1964).
92. Richardson, E. *et al.* The receiver operating characteristic curve accurately assesses imbalanced datasets. *Patterns* **5**, 100994 (2024).
93. Turck, N. *et al.* pROC: an open-source package for R and S+ to analyze and compare ROC curves. *BMC Bioinformatics* **8**, 12–77 (2011).
94. DeLong, E. R., DeLong, D. M. & Clarke-Pearson, D. L. Comparing the areas under two or more correlated receiver operating characteristic curves: a nonparametric approach. *Biometrics* **44**, 837–845 (1988).